\documentclass[twocolumn,nofootinbib]{revtex4-1}
\usepackage[utf8]{inputenc}
\usepackage{amsmath}
\usepackage{graphicx}
\usepackage{subcaption}
\usepackage{amssymb}
\usepackage{color,colortbl}
\usepackage{cancel}
\usepackage{framed}
\usepackage{comment}
\usepackage[margin=0.7in]{geometry}
\usepackage{mathtools}
\usepackage{hyperref}
\usepackage{chngcntr}
\usepackage{soul}
\definecolor{Gray}{gray}{0.9}
\definecolor{LightCyan}{rgb}{0.88,1,1}

\counterwithin{equation}{section}

\makeatletter

    \def\CT@@do@color{%
      \global\let\CT@do@color\relax
            \@tempdima\wd\z@
            \advance\@tempdima\@tempdimb
            \advance\@tempdima\@tempdimc
    \advance\@tempdimb\tabcolsep
    \advance\@tempdimc\tabcolsep
    \advance\@tempdima2\tabcolsep
            \kern-\@tempdimb
            \leaders\vrule
                    \hskip\@tempdima\@plus  1fill
            \kern-\@tempdimc
            \hskip-\wd\z@ \@plus -1fill }
    \makeatother

\newcommand{\pd}{\partial}

\newcommand{\nn}{\nonumber}

   \def\e{\epsilon}
\def\d{\delta}
 \def\m{\mu}   
   \def\l{\lambda}
\def\D{\Delta}

 \newcommand{\Ocal}{{\mathcal O}}

\newcommand*{\affaddr}[1]{#1} % No op here. Customize it for different styles.
\newcommand*{\affmark}[1][*]{\textsuperscript{#1}}

\def\Cincy{\small{Department of Physics, University of Cincinnati, Cincinnati, Ohio 45221, USA}}
\def\Weizmann{\small{Department of Particle Physics and Astrophysics, Weizmann Institute of Science, Rehovot 761001, Israel}}
\def\Harvard{\small{Department of Physics, Harvard University, 17 Oxford St., Cambridge, MA 02138, USA}}

\begin{document}

% Don't want date printed
\date{\today}
% Make title large and bold
\title{\Large\bfseries %Dilute and dense axion stars
	Global view of QCD axion stars}

% Target typesetting:
%
% Author A, Author B, Author C, Author D and Author E
%        A,B,C Department of Computer Science
%       D,E Department of Mechanical Engineering
%          Email A,B,C,D,E @university.edu
%                  Latex University

\author{
Joshua Eby,\affmark[$1$] 
Madelyn Leembruggen,\affmark[$2$] 
Lauren Street,\affmark[$3$] 
Peter Suranyi,\affmark[$3$] 
and L.C.R.Wijewardhana\affmark[$3$] \\
{\it\affaddr{\affmark[$1$]\Weizmann}} \\
{\it\affaddr{\affmark[$2$]\Harvard}} \\
{\it\affaddr{\affmark[$3$]\Cincy}}
}

\begin{abstract}
Taking a comprehensive view, including a full range of boundary conditions, we reexamine QCD axion star solutions based on the relativistic Klein-Gordon equation (using the Ruffini-Bonazzola approach) and its non-relativistic limit, the Gross-Pitaevskii equation.  A single free parameter, conveniently chosen as the central value of the wavefunction of the axion star, or alternatively the chemical potential with range $-m<\mu< 0$ (where $m$ is the axion mass), uniquely determines a spherically-symmetric ground state solution, the axion condensate.  We clarify how the interplay of various terms of the Klein-Gordon equation determines the properties of solutions in three separate regions: the structurally stable (corresponding to a local energy minimum) dilute and dense regions, and the intermediate, structurally unstable transition region. From the Klein-Gordon equation, one can derive alternative equations of motion including the Gross-Pitaevskii and Sine-Gordon equations, which have been used previously to describe axion stars in the dense region.  In this work, we clarify precisely how and why such methods break down as the binding energy increases, emphasizing the necessity of using the full relativistic Klein-Gordon approach.
%We point out how axion number violating axion star decay processes affect the viability of those solutions, as dark matter.  
Finally, we point out that, even after including perturbative axion number violating corrections, solutions to the equations of motion, which assume approximate conservation of axion number, break down completely in the regime with strong binding energy, where the magnitude of the chemical potential approaches the axion mass.
\end{abstract}

\maketitle

\section{Introduction}

Gravitationally bound states of scalar excitations, termed boson stars, have been studied extensively over the past half century. Investigation of scalar boson stars started with the analysis of the works of Kaup~\cite{Kaup}, and Ruffini and Bonazzola (hereafter, RB) \cite{RB} (and more recently using the same method,~\cite{Breit}).  They identified a maximum mass for boson stars consisting of non-interacting bosons, above which they become structurally unstable\footnote{We adopt the terminology of \emph{structural} (in)stability, as opposed to the more typical \emph{gravitational} (in)stability, in order to emphasize that this is driven by self-interactions as much as by gravitational interactions.  Structural stability is the requirement that the solution is at a minimum of the energy functional, as explained in~\cite{GRB}.}.  Later, Colpi et al.~\cite{Colpi} derived a maximum mass for boson stars with repulsive interactions. Various aspects of self-gravitating objects in astrophysics and cosmology have also been investigated~\cite{Friedberg,SS,SS2,Liddel,Lee,Khlopov1,Khlopov2,Khlopov3}. 

A recent surge in studies of boson stars~\cite{SikivieYang,BB1,ESVW,GHPW,BraatenDense,ChavanisCollapse,ELSW,Davidson,Collisions,ELSW2,WilczekASt,Hertzberg,Hertzberg2,Hertzberg3,Ansatz,Sarkar} stems, in part, from the renewed interest in determining whether dark matter (DM) could consist of condensates of axions or other axion like particles. A particularly well-motivated scalar DM candidate is the QCD axion, parametrized by a decay constant $f= 6\times10^{11}$ GeV and particle mass $m=10^{-5}$ eV; as a result, bound states of QCD axions (which we will call QCD axion stars) have received special attention \footnote{Both $m$ and $f$ can shift by a few orders of magnitude without violating any experimental constraint; however, for QCD axions the product $m\,f$ is fixed. We use the values quoted above as a benchmark for parameter estimation.}. Axion stars were analyzed by Barranco and Bernal (BB)~\cite{BB1} using the formalism employed by RB, and in doing so these authors derived the relevant Einstein-Klein-Gordon (EKG) equations describing axion stars. In this study, we will refer to this formalim as the EKG formalism.  This was a unique enterprise because of the leading-order attractive interactions of axions, which was not previously taken into account \cite{Kaup,RB}. They looked for solutions in two regions of parameter space: first, where the axion decay constant is very large, approaching the Planck mass $M_P = 1/\sqrt{G}$ where $G$ is the gravitational constant; and second, where the mass and decay constant are those of QCD axions. In the former range ($f$ close to $M_P$), they identified a maximum mass for axion stars. In the second parameter range (for QCD axions), they found a handful of solutions where the masses and radii of the axion stars are in the range $10^{14}$ kg and a few meters (respectively), which were the first known QCD axion star solutions.

However, the scaling relations used in \cite{BB1}, which worked well when $f$ was near $M_P$, made solutions to the equations of motion difficult to find for QCD axion parameters. As a result, BB did not find {\it dilute} structurally stable QCD axion stars, or their maximum mass. Nearly a decade later, Chavanis and Delfini \cite{ChavanisMR,ChavanisMR2} analyzed boson star configurations with self-interactions in the nonrelativistic limit using the Gross-Pitaevskii (GP) equation, and derived a general bound on the mass of attractively-interacting boson condensates as a function of the 4-point coupling $\l$. To investigate dilute axion star solutions using the EKG formalism, the key is the rescaling of the relativistic field and the radial coordinate using the scaling parameter $\Delta= \sqrt{1-\e^2}$, where $m\,\e$ is the energy eigenvalue of the axion, related to the the chemical potential via $\mu=m(\e-1)$.  For QCD axions, the EKG formalism was applied with appropriate scaling relations to determine the maximum mass, $M_{\rm max} = 10M_P f/m \approx 10^{19}$ kg \cite{ESVW}. In this dilute branch of solutions the radius scales inversely as the mass \cite{ChavanisMR,ChavanisMR2,ESVW,GHPW,Ansatz}. \footnote{In \cite{ESVW}, though the numerical solutions were correct, the structurally stable/unstable branches of solutions were misidentified; in fact, the more dilute solutions where $M \propto R^{-1}$ are stable, whereas the other branch with $M\propto R$ are unstable.}

The BB solutions for QCD axion stars had masses much lower than the maximum. 
%For the same values of $f$ and $m$ relevant for QCD axions, a more dilute branch of bound states exist, where the maximum mass of $10^{19}$ kg corresponds to a size of a few hundred kilometers \cite{ESVW,GHPW}. In this dilute branch of solutions the radius scales inversely as the mass \cite{ChavanisMR,ChavanisMR2,ESVW,GHPW,Ansatz}
%, and therefore a solution in the mass range of $10^{14}$ kg in the dilute branch would have a size $10^5$ times larger than the size of the solution with maximum mass, i.e. $\Ocal(10^7)$ km. 
It is now understood that in this mass range, there are as many as three solutions to the equations of motion: a \emph{dilute} solution with radius $\Ocal(10^7)$ km; a \emph{transition} solution with radius $\Ocal(10)$ meters; and a \emph{dense} solution with radius as small as $\Ocal(10)$ cm. The solutions found by BB for QCD axion stars fall in the range of transition axion star solutions which, as it turns out, are structurally unstable to collapse, as they correspond to a maximum rather than a minimum of the energy functional. Collapsing axion stars evaporate a large fraction of their mass through rapid emission of relativistic axions \cite{ELSW,ELSW2,Levkov,Helfer,Michel}.

Dense axions stars were proposed by Braaten at al. \cite{BraatenDense}, who used the nonrelativistic GP formalism. However, on the dense branch of solutions, it is now understood that relativistic corrections become large \cite{WilczekASt,GRB}. In this work, we will clarify the range of applicability of the GP formalism both by analysis of its derivation and by direct comparison to the KG equation. Several other methods have been proposed to describe axion stars in this regime, including the Sine-Gordon (SG) equation, and we will clarify the applicability of these methods as well.

At the crossover from transition to dense solutions, there is in fact a minimum mass which is about an order of magnitude smaller than the mass of the lighest BB solution, which is calculable using the EKG equations as well. We will also point out that the EKG equations in fact break down at extremely large $\D \sim \mathcal{O}(1)$, corresponding to increasingly massive states on the dense branch. This suggests a gap in the current understanding of the dense branch of axion stars, as all known methods break down in the regime of large relativistic corrections.

This paper is organized as follows: In Section \ref{sec:Relativistic}, we describe in detail the calculation of axion star properties using the EKG equations, comparing the contributions of different terms in the calculation of the total mass and analyzing where this method becomes inadequate; in Section \ref{sec:OtherMethods}, we show how alternative methods used in the literature can be derived from the EKG equations, and compare the results to see where they break down.  We conclude in Section \ref{sec:Conclusions}.

We will use natural units throughout, where $\hbar = c = 1$.
 
 \section{Relativistic theory} \label{sec:Relativistic}

\subsection{Einstein-Klein-Gordon (EKG) Equations of Motion} \label{sec:EKG}

In this section we review the basic equations of motion describing axion stars. The focus of this work is an axion theory defined by a scalar field $\phi$ with potential
\begin{equation} \label{Vphi}
 V(\phi) = m^2\,f^2\,\left[1 - \cos\left(\frac{\phi}{f}\right)\right].
\end{equation}
We focus on this potential because it is relevant for QCD axions as well as many classes of axion-like particles arising from broken global symmetries in the early universe.  A more general analysis might allow the coefficients of the self-interaction terms to vary in sign or magnitude, an interesting case that deserves separate treatment.

The field is also coupled to gravity, so the resulting equations of motion are the EKG equations with the gravitational metric
\begin{equation}
 ds^2 = -B(r)\,dt^2 + A(r)\,dr^2 + r^2\,d\Omega^2,
\end{equation}
where we have assumed spherical symmetry. For a scalar field condensate, one can evaluate the EKG equations as an expectation value in $N$-particle states (as described by \cite{RB}), expanding $\phi$ in ground state creation and annihilation operators $a_0^\dagger$ and $a_0$ as
\begin{equation} \label{phiRB}
 \phi(t,r) = R(r) \left[a_0\,e^{-i\,\e\,m\,t} + a_0^\dagger\,e^{i\,\e\,m\,t}\right],
\end{equation}
where the wavefunction $R(r)$ has a ground state eigenenergy $\e\,m < m$ (the quantity $\mu = m(\e-1)$ is the chemical potential of axions in the condensate).  The limitations of this method, pioneered by RB, will be described in more detail in Section \ref{sec:GRB}.  The resulting EKG equations of motion are
\begin{widetext}
\begin{align} \label{EKG}
\frac{A'(y)}{A(y)}&=\frac{1-A(y)}{y} + 2\pi\,\delta\,y\,A(y)\left\{\frac{\epsilon{}^2\,Z(y)^2}{B(y)}+\frac{Z'(y)^2}{A(y)}+4[1-J_0(Z(y))]\right\},\nn\\
\frac{B'(y)}{B(y)}&=\frac{A(y)-1}{y} + 2\pi\,\delta\,y\,A(y)\left\{\frac{\epsilon{}^2 \, Z(y)^2}{B(y)}+\frac{Z'(y)^2}{A(y)}-4[1-J_0(Z(y))]\right\},\nn\\
Z''(y)&=-\left[\frac{2}{y}+\frac{B'(y)}{2\,B(y)}-\frac{A'(y)}{2\,A(y)}\right]Z'(y)-A(y)\left[\frac{\epsilon{}^2\,Z(y)}{B(y)}-2\,J_1(Z(y))\right],
\end{align}
\end{widetext}
where we have introduced the rescaled variables $Z(y) = 2\sqrt{N}R(r)/f$, $y=m\,r$, and $\d = f^2/M_P^2$. Eqs. (\ref{EKG}) have also been referred to as the RB equations in the literature \cite{BB1,ESVW}.  Whenever numerical values of physical parameters are presented in this paper we will use the value $\delta = 2.5\times 10^{-15}$, a typical value for QCD axions.
Note for future reference that the original $\cos(\phi/f)$ potential transforms to a Bessel function $J_0(Z)$ when expectation values of the equations of motion are taken \cite{ESVW}.

We solved Eqs. (\ref{EKG}) by imposing the following set of conditions on $A(y)$, $B(y)$, and $Z(y)$:  
\begin{itemize}
 \item[1.] $A(0)=1$;
 \item[2.] $Z(y)$ is regular and finite at $y=0$;
 \item[3.] $Z(y)$ approaches zero at some $y_{\rm max}$ with arbitrary precision;
 \item[4.] $A(y_{\rm max})B(y_{\rm max})=1$, implying that the metric turns Schwarzschild  ``outside"  the star.
\end{itemize}
The point at which the wavefunction approaches zero determines the radius $R_{99}$ of the star (inside which $0.99$ of the mass of the star is contained), which is a single free parameter characterizing the family of solutions. The radius has a one-to-one relationship with the central density $Z(0)^2$, which (following the usual convention) we take to be the input parameter to our numerical calculations. Alternatively, the system could be solved by first fixing $\e$, which also has a one-to-one monotonic relationship with $Z(0)$. 
%At every $Z(0)$ one can easily solve the system of equations numerically, though the solution becomes increasingly difficult at $Z(0)\gtrsim 100$.  However,  as we will be explain later, solutions based on the system of Eqs. (\ref{EKG}) are  unphysical at $Z(0)\gtrsim 10$.

At every value of $Z(0)$ we find a {\em unique}, spherically symmetric, nodeless solution for the wavefunction $Z(y)$. 
%The physical properties of solutions to the EKG equations  depend critically on the value of  $Z(0)$. One important parameter determining the  physical properties of solutions is $\delta=f^2\,/\,M_P{}^2$.  
Solutions can be divided  into three branches based on the central field value:
\begin{itemize}
 \item Dilute: $Z(0) < Z_{\rm dilute} \approx 6 \sqrt{\delta}$;
 \item Transition: $Z_{\rm dilute}<Z(0)<Z_{\rm dense}$;
 \item Dense: $Z(0) > Z_{\rm dense} \simeq 3.5$
\end{itemize}
 which we describe below.
 
 The energy eigenvalue $\e\,m$ has a one to one correspondence with $Z(0)$ as well. When $1 - \e \ll 1$, the field is very nonrelativistic, but as we shall see, in the crossover region and on the dense branch of solutions this condition is no longer satisfied. To quantify the breakdown of the nonrelativistic approximation, we define the following approximate regions of parameter space:
\begin{itemize}
 \item Nonrelativistic: $\e \gtrsim 0.9$;
 \item Quasi-Relativistic: $0.5 \lesssim \e \lesssim 0.9$;
 \item Ultra-Relativistic: $\e \lesssim 0.5$.
\end{itemize}
We will discuss these conditions in what follows.

\subsection{Solutions}
Starting at the lowest values of $Z(0)$, solutions for $Z(0) \leq Z_{\rm dilute} \approx 6 \sqrt{\delta}$ belong to the structurally stable  \emph{dilute} branch of solutions \cite{ChavanisMR,ESVW}. On this branch, gravity is Newtonian and it is sufficient to take only the leading-order self-interaction term, which is attractive and proportional to $\phi^4 \propto Z^4$. In this regime, direct numerical solutions of the system (\ref{EKG}) become more and more difficult, as the magnitude of the chemical potential approaches zero ($m|1-\e| \ll m$), the radius of the star becomes large ($R_{99} \gg 1/m$), and the field becomes weak ($\phi \ll f$). 

The most efficient method to find numerical solutions on the dilute branch is to rescale all physical variables using the scale parameter $\Delta=\sqrt{1-\e{}^2}$, by introducing the new radial coordinate and field 
\begin{equation}
 x= \Delta\, y, \qquad Z(y)=\Delta\, Y(x).
\end{equation}
Then, a systematic expansion of Eqs. \eqref{EKG} as a power series in both $\delta$ and $\Delta$ give rise to a much simpler set of coupled equations which are appropriate in the dilute region \cite{ESVW},
\begin{align} \label{rescale}
 a'(x) &= \frac{x}{2}\,Y(x)^2 - \frac{a(x)}{x}, \,\,\qquad b'(x) = \frac{a(x)}{x}, \nn  \\
 Y''(x)&= - \frac{2}{x}Y'(x)-\frac{1}{8}\, Y(x)^3 
	  + [1 + \kappa\,b(x)]\,Y(x).
\end{align}
where the metric functions have been expanded using $A(r) = 1 + \delta\,a(x)$ and $B(r) = 1 + \delta\,b(x)$.  The effective coupling of the field $Y(x)$ to gravity is given by $\kappa \equiv 8 \pi \delta / \Delta^2$. These equations are exactly equivalent \cite{Ansatz} to the nonrelativistic Gross-Pitaevskii-Poisson (GPP) equations, which we will discuss in Section \ref{sec:GPP}, and the solutions have been discussed many times in the literature \cite{ChavanisMR,ChavanisMR2,ESVW,Hertzberg,WilczekASt}.

\begin{figure}[t]
\centering 
\includegraphics[scale=0.50]{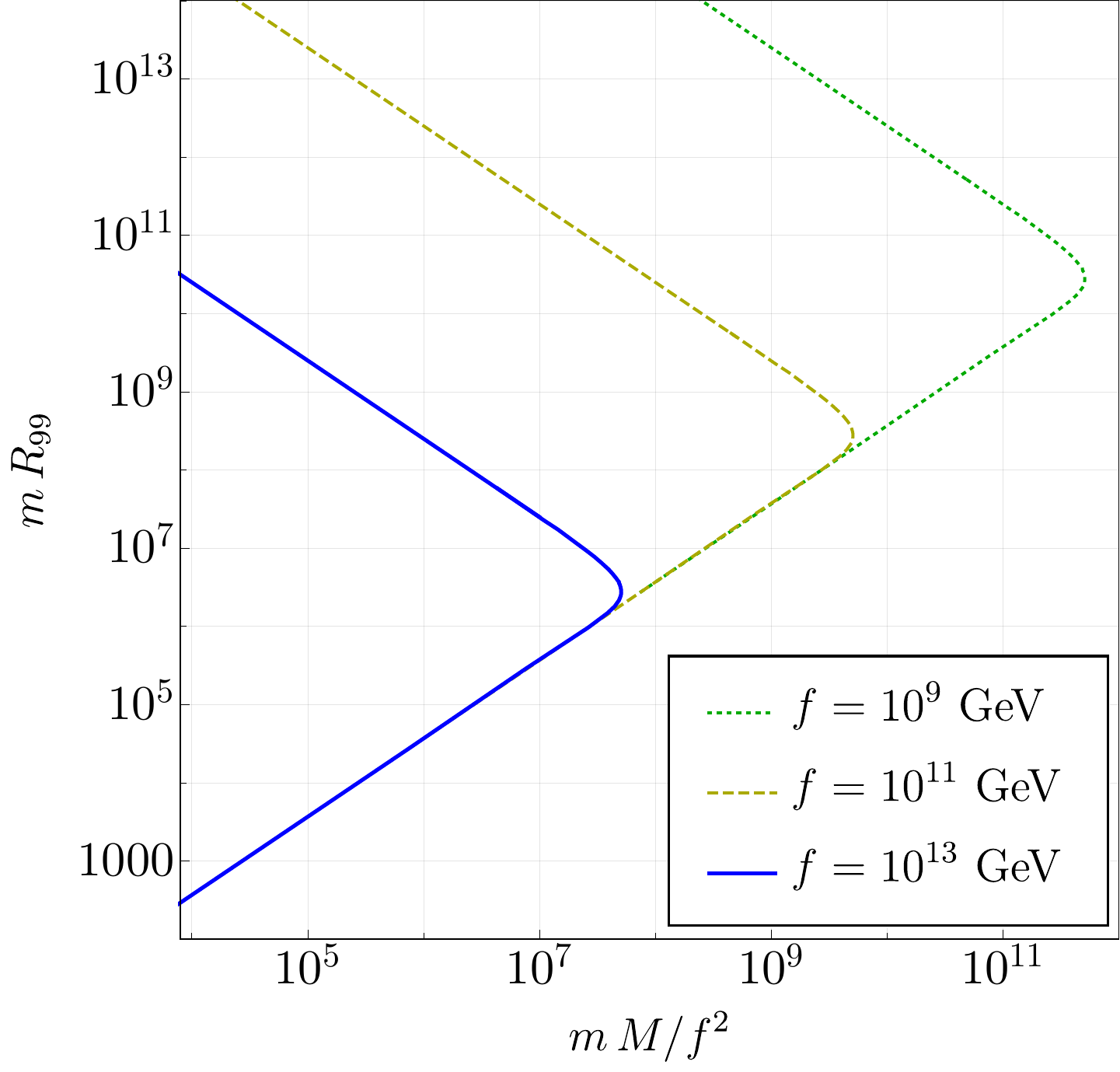}
\,
\includegraphics[scale=0.255]{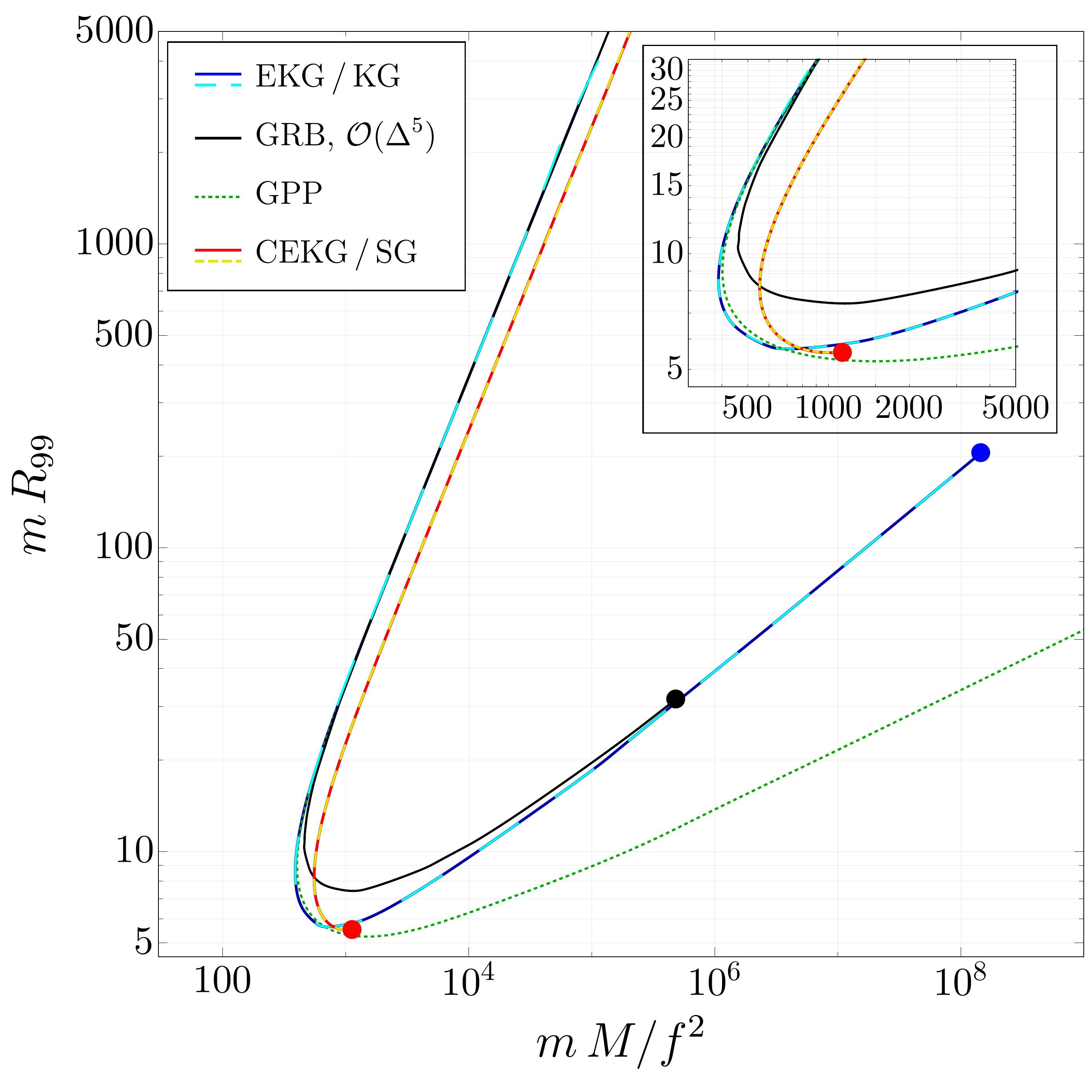}
\caption{Top:  Mass $M$ and radius $R_{99}$ of axion stars in the dilute region for different choices of axion decay constant $f$ using the EKG method (Eq. (\ref{rescale})). The maximum mass at $M_{\rm max} = 10\,M_P\,f/m$ marks the crossover from the dilute (upper, downward-sloping) to transition (lower, upward-sloping) branches of solutions.  Bottom: The masses and radii of axion stars in the vicinity of the dense crossover for the different methods analyzed in this study; inset shows the crossover from transition to dense branches of solutions.  The blue, black, and red dots mark the endpoint for solutions of the EKG, GRB, and CEKG methods (respectively), as described in the text.} \label{fig:RvsM_dilute_methods}
\end{figure}

For completeness, we reproduce the well-known mass-radius relation for axion stars in the dilute region in the top panel of Figure \ref{fig:RvsM_dilute_methods} for different decay constants, $f$.  We choose decay constants in the allowed range for QCD axions where the dashed line approximately corresponds to the value we use in this study, $f = 6 \times 10^{11} \, \text{GeV}$.  The top panel of Figure \ref{fig:RvsM_dilute_methods} illustrates clearly the existence of a maximum mass ${M_{\rm max} \simeq 10 \,M_P\,f/m}$, which occurs at $Z(0) =Z_{\rm dilute}$.  The large-radius curves away from the maximum mass constitute the \emph{dilute} axion stars; the smaller-radius curves constitute the \emph{transition} branch which will be discussed later in this section.  Note that axion stars with extremely small central density $Z(0)$, corresponding to the ultra-dilute region near the top of the upper panel of Figure \ref{fig:RvsM_dilute_methods}, are very well described by a non-interacting approximation; self-interactions become important only near $M_\text{max}$ and at larger $Z(0)$.

Dilute axion stars are fully stable, both against decay \cite{ESW,MTY,BraatenEmission,EMSW} as well as under perturbations (what we call \emph{structural stability}) \cite{LeeStability,Mielke,Toth,ChavanisMR}. As such, their phenomenological effects can be searched for in the dark matter halo. Searches for effects of dilute axion stars include: collisions with neutron stars giving rise to high-intensity radio photon emission \cite{Iwazaki,TkachevFRB}; microlensing \cite{MarshMC}; transient effects from rare encounters of an axion star with Earth \cite{Collisions,GNOME}; or possible capture in the solar system leading to high-density subhalos \cite{RHalo}. This field continues to attract increasing interest and new ideas for how to probe dilute axion stars in the halo. In this work, however, we will concentrate on the region of larger $Z(0)$, since our goal is to clarify the status of more dense configurations, and there is little controversy about the properties of dilute axion stars in the recent literature.

\begin{figure}
\centering
	\begin{subfigure}[t]{0.388\textwidth}
	\includegraphics[width=\textwidth]{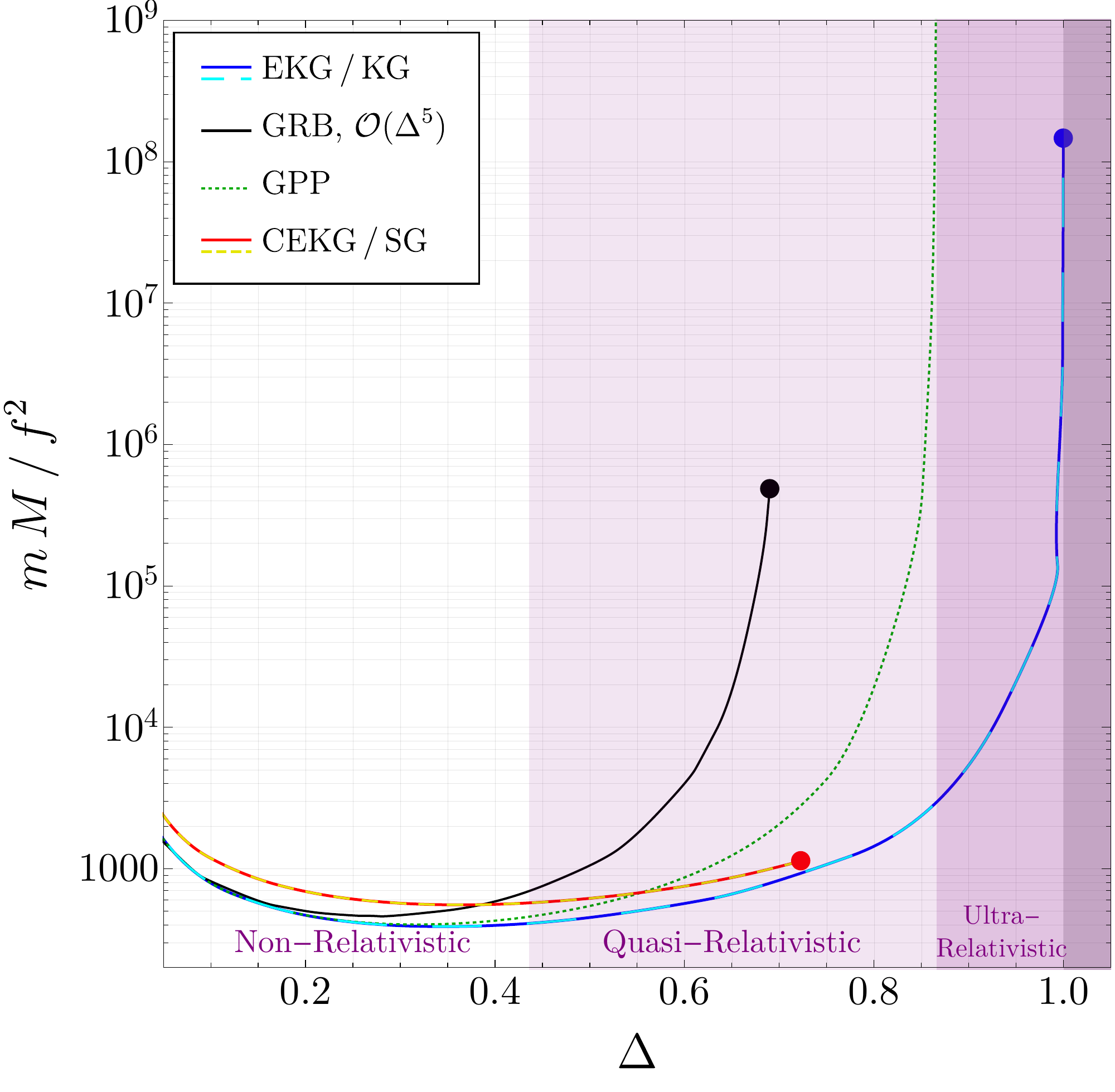}
	\end{subfigure}
	\,\,\,\,\,
	\begin{subfigure}[t]{0.388\textwidth}
	\includegraphics[width=\textwidth]{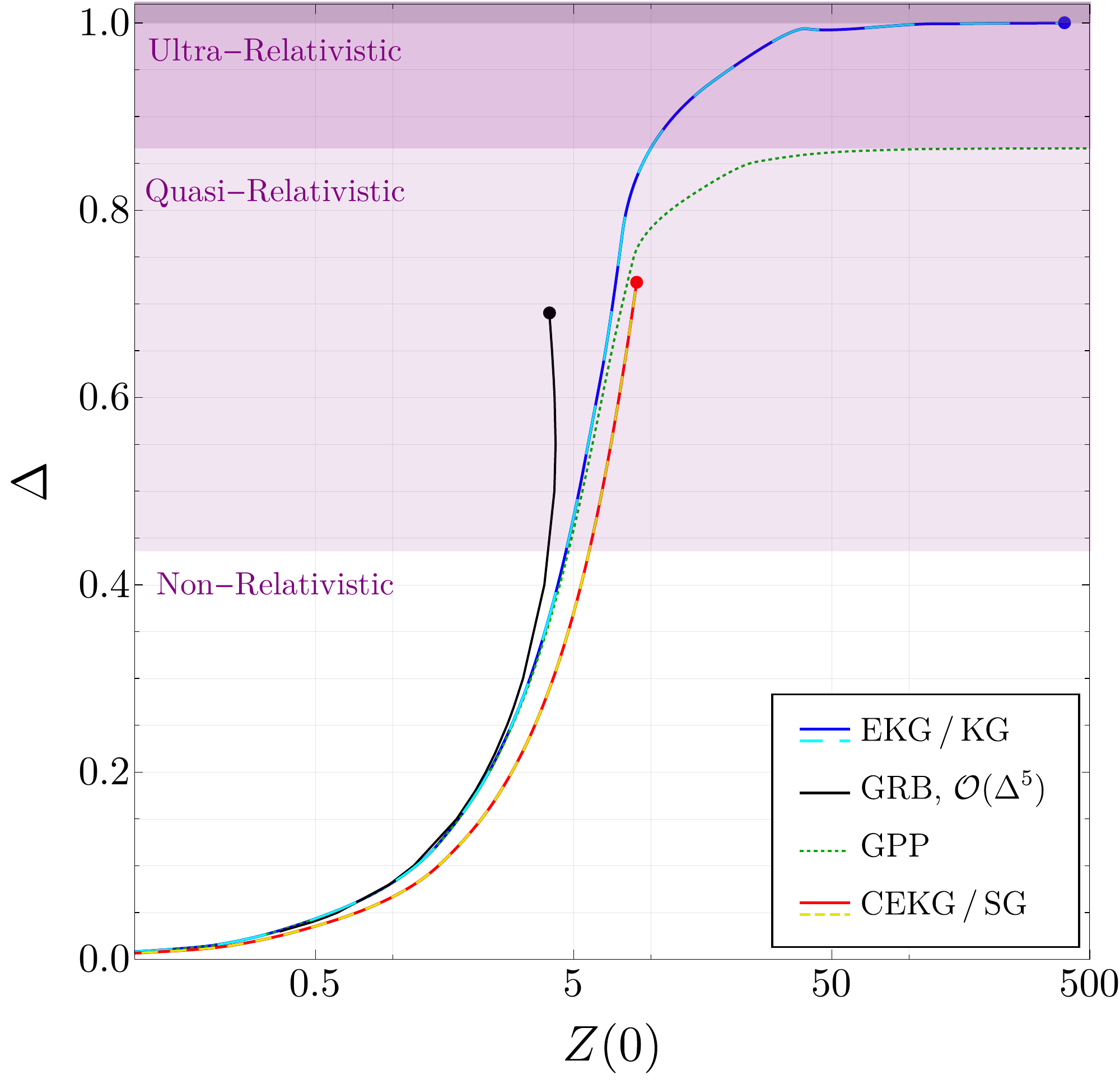}
	\end{subfigure}
	~
	\begin{subfigure}[t]{0.408\textwidth}
	\includegraphics[width=\textwidth]{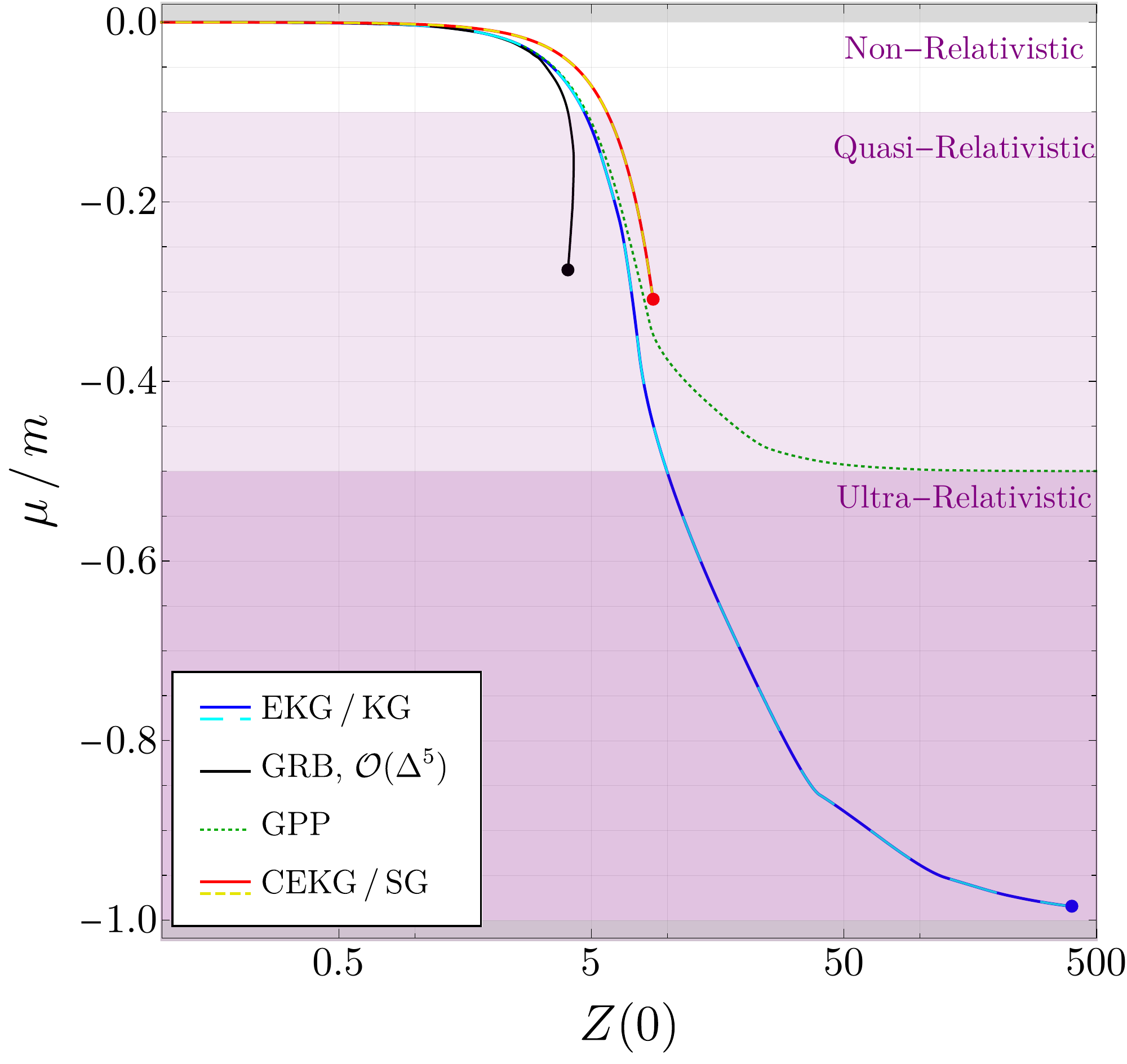}
	\end{subfigure}
	\,
\caption{Total axion star mass as a function of $\D$ (top), as well as the binding energy parameter $\D$ (middle) and chemical potential per unit mass $\m/m$ (bottom) as a function of central field value $Z(0)$ for different methods. In all panels, the light purple (dark purple) shaded region represents the quasi-relativistic (ultra-relativistic) region defined in Section \ref{sec:EKG}.} \label{fig:methods}
\end{figure}

The results of our numerical calculations away from the dilute region, $Z(0)\geq 0.01$, using the EKG formalism of Eq. (\ref{EKG}) are tabulated in Table \ref{data} and depicted as the dark blue line in the bottom panel of Figure \ref{fig:RvsM_dilute_methods} and in all panels of Figure \ref{fig:methods}.  The different methods depicted in the bottom panel of Figure \ref{fig:RvsM_dilute_methods} and in all panels of Figure \ref{fig:methods} will be discussed in more detail in Section \ref{sec:OtherMethods}.  Solutions with $Z_{\rm dilute}<Z(0)<Z_{\rm dense}$ correspond to the  \emph{transition} branch, and are structurally unstable (gray entries in the table). Note in particular that, for $0.1 \lesssim Z(0) \lesssim 1$, we reproduce roughly the original BB solutions \cite{BB1}; the slight deviations in the numerical results are due to the fact that BB truncated the self-interaction potential at $\Ocal(Z^6)$, whereas we used the full potential. Thus the BB solutions lie on the transition, not dilute, branch of axion star solutions. 

At larger values of $Z(0)$, the transition branch crosses over to a \emph{dense} branch. From the bottom panel of Figure \ref{fig:RvsM_dilute_methods}, one can see that at the crossover point $Z_{\rm dense} \simeq 3.5$, there is a minimum value of the axion star mass $M_{\rm min}\simeq 39\,\sqrt{\delta}\,M_{\rm max} \simeq 390 f^2 / m$; for our benchmark QCD parameters, this gives $M_{\rm min} \approx 2\times 10^{13}$ kg.  Solutions at $Z(0) > Z_\text{dense}$ along the dense branch and at $Z(0) < Z_\text{dilute}$ on the dilute branch are structurally stable, while solutions on the transition branch, $Z_\text{dense} > Z(0) > Z_\text{dilute}$ are structurally unstable.

At very large $Z(0)\gtrsim 100$, the solutions become increasingly sensitive to the input boundary conditions, and as a result become extremely hard to calculate.  We represent the cutoff of our numerical solutions at $Z(0) \simeq 400$ by the blue dot in the lower panel of Figure \ref{fig:RvsM_dilute_methods}. However, solutions with extremely large $Z(0)\gtrsim 10$ are unphysical, because they exist in an ultra-relativistic domain where $\epsilon\ll 1$, as illustrated in the middle panel of Figure \ref{fig:methods}. In this region, the binding energy per particle exceeds $m/2$, giving rise to a large negative chemical potential (bottom panel of Figure \ref{fig:methods}); this implies that a large number of axions can easily pop in and out of the condensate from the vacuum, a phenomenon which in a field theory usually necessitates the introduction of  {\em relativistic} loop corrections. We refer to Section \ref{sec:OtherMethods} for further details of these calculations.

Such unphysical solutions are surely an artifact of the breakdown of the equations of motion describing the system. Indeed, the formalism we use is based on an approximation which assumes the conservation of axion number.  Even if we add  the contribution of higher harmonics \cite{GRB} (described in Section \ref{sec:GRB}) which violate axion number conservation, it only provides a perturbative improvement of the axion number conserving theory. It is likely that by including higher harmonic corrections, we merely include tree diagrams in a yet-unknown field theory of relativistic axion condensates. In Section \ref{sec:OtherMethods}, we will describe alternative methods for describing axion stars in this regime, emphasizing how and why each method breaks down.

Finally, though states above the dense minimum,  $Z_{\rm dense}\simeq 3.5$, are structurally stable, it  is important to remember that all condensates with central field value $Z(0)\gtrsim 0.05$, produced after the big bang would have decayed completely by the present time and cannot form even a fraction of dark matter \cite{ESW,EMSW}.

\subsection{Relative Magnitudes of Energy Terms}
Here, we discuss the contributions of the kinetic and gravitational energy terms to the total mass of the condensate, which is defined by the volume integral over the $tt$ component of the stress-energy tensor, 
\begin{align} \label{densepotential}
 M &= \int d^3r \sqrt{-g}\, T_{00}  \nn \\
 		&= \frac{ f^2}{4 \,m}\int d^3y \,\sqrt{A B}\left[\frac{\e^2\,Z^2}{B} + \frac{Z'^2}{A} + 4\left(1-J_0(Z)\right)\right].
\end{align}
In this work, we forgo comparison of the contribution from the self-interaction energy term, for reasons that will be discussed below.

In the extremely nonrelativistic region where $Z(0)\lesssim 0.01$, the expansion of the Bessel function in powers of $Z$ converges fast and the potential $4 (1 - J_0(Z))$ can be written as $Z^2+ V(Z)$, where the magnitude of the self-interaction, $|V(Z)| \ll Z^2$.  In this case, the expression of the total mass is dominated by the term
 \begin{equation}\label{massindilute}
 M\simeq\int d^3 r \, T_{00}\simeq\frac{f^2}{2\,m}\int d^3y \,Z^2.
 \end{equation}
While that term provides the normalization of the wave function, it plays no direct role in finding the numerical solution. More precisely, only the other terms (kinetic, gravitational, and self interaction) are relevant for the determination of the properties of the solution. Therefore it is meaningful to calculate the relative contribution of those three terms to either Eq. (\ref{massindilute}) or to the total energy density, as was done in \cite{BraatenReview}.
  
In the dense region, where $Z(0)=\Ocal(1)$ or larger, the term $Z^2$ no longer dominates the potential and the total mass must be calculated using the full energy-momentum tensor, Eq. (\ref{densepotential}).  Attempting to separate the self-interaction energy from the third term in Eq. (\ref{densepotential}) no longer adds anything to the description of the system and the contribution of the self-interaction energy to the total mass is not as apparent as in the nonrelativistic region where the dominant quartic interaction term also scales with $\Delta^2$.  Therefore, in this work we focus only on the contributions from the gravitational and kinetic terms by taking ratios of these terms to Eq. ({\ref{densepotential}}), where the gravitational contribution is scaled by $\delta$ and the kinetic contribution by $\Delta^2$.

\begin{table*}[t]
\centering
%\hspace{-50pt}
\noindent%\(\pmb{
$\begin{array}{| c | c | c | c | c | c | c | c | c |}
\hline
 Z(0) & M\,m/f^2 & M \text{[kg]} & m \,R_{99} & R_{99} \text{ [m] } & \Delta  & \epsilon  & K/M& M_g/M\\
\hline \hline
\rowcolor{LightCyan}
100 & 2.18\times 10^6&2.29\times 10^{18} &50.32&8.3& 0.998&0.0612&0.367&-5.3\times 10^{-12}\\
\hline
\rowcolor{LightCyan}
50 &2.68\times 10^5& 1.71\times 10^{16}& 25.5& 4.2& 0.9926 & 0.1213&0.364& -1.1\times 10^{-12}\\
\hline
 10 & 2972 & 1.9\times 10^{14} & 6.96 & 0.136 & 0.8666 & 0.499 & 0.317&5\times 10^{-14}\\
\hline
 8 & 1272 & 6.14\times 10^{13} & 5.88 & 0.115 & 0.7774 & 0.6289 & 0.311&3.8\times 10^{-14} \\
\hline
 7 & 779 & 5\times 10^{13} & 5.66 & 0.112 & 0.6876  & 0.7261 &0.287&5\times 10^{-14}\\
\hline
 6 & 540.5 & 3.46\times 10^{13} & 5.96 & 0.116 & 0.5816 & 0.8135 &0.245&6.6\times 10^{-14}\\
\hline
 5 & 429.1 & 2.75\times 10^{13} & 6.685 & 0.13 & 0.4705 & 0.8824 &0.192&7.9\times 10^{-14} \\
\hline
 4 & 390.1 & 2.5\times 10^{13} & 8.22 & 0.16 & 0.3623 & 0.9321 &0.135&8.9\times 10^{-14}\\
\hline
\rowcolor{Gray}
 3 & 417 & 2.67\times 10^{13} & 10.96 & 0.214 & 0.2611 & 0.9611 &0.082 &1\times 10^{-14}\\
\hline
\rowcolor{Gray}
 2 & 525.1 & 3.36\times 10^{13} & 19.55 & 0.38 & 0.1682 & 0.98575 &0.038&1.02\times 10^{-13}\\
\hline
\rowcolor{Gray}
 1 & 956.5 & 6.12\times 10^{13} & 33.78 & 0.66 & 0.0822 & 0.9966 &0.0099&1.02\times 10^{-13}\\
 \hline
 \rowcolor{Gray}
 0.3&3109&1.99\times 10^{14}&116.1&2.2&0.0245&0.997&10^{-3}&1.02\times 10^{-13}\\
\hline
\rowcolor{Gray}
 0.1 & 9276 & 5.94\times 10^{14} & 338.2 & 6.6 & 0.00815 & \sim 1& 10^{-4}&1.02\times 10^{-13}\\
\hline
\rowcolor{Gray}
 0.01 & 92797 & 5.94\times 10^{15} & 3401 & 66 & 0.000815 & \sim 1& 10^{-6}&1.12
 \times 10^{-13}\\
\hline
\end{array}$
\caption{Numerical results for central field value $Z(0)$, total mass (in scaled and physical units), radius (in scaled and physical units), binding energy parameter $\D$, scaled energy eigenvalue $\e$, and relative contributions $K/M$ and $M_g/M$ to the kinetic and gravitational energies, respectively. All were obtained by solving the EKG equations (\ref{EKG}) in the dense region, including the dense minimal mass. The gray background signifies structurally unstable solutions, while the cyan background signifies unphysical solutions in the ultra-relativistic region, where the chemical potential $\m \approx 0$.  Unshaded rows correspond to structurally stable but present-day unviable configurations.}
\label{data}
\end{table*}

\begin{figure}[t]
\centering 
\includegraphics[scale=0.4]{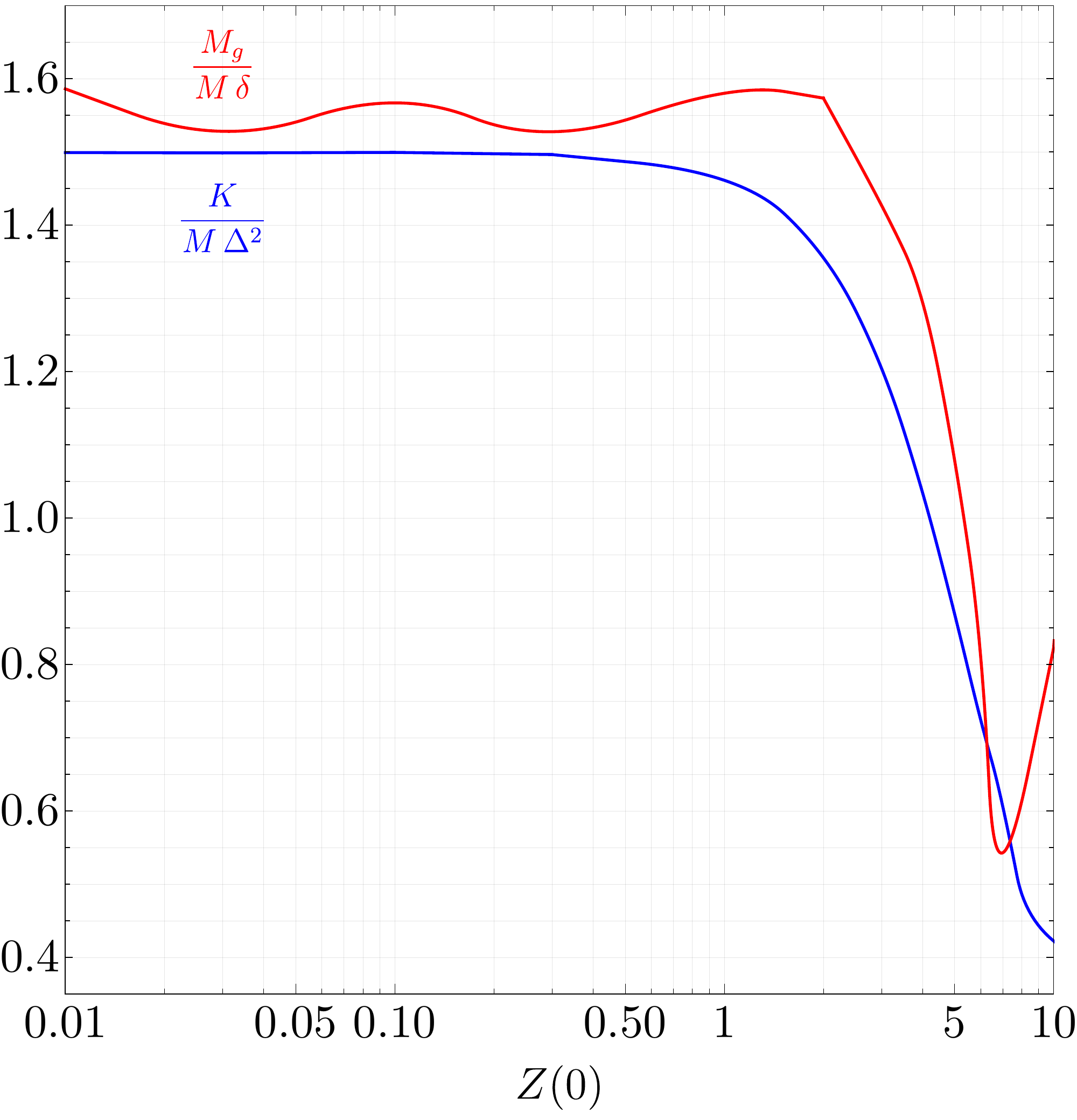}
\caption{Plot of $K\,/(M\,\Delta^2)$ (blue) and $M_g\,/(M\,\delta)$ (red) as a function of Z(0).}\label{KandGplot}  
\end{figure}

Table \ref{data} shows the contribution of the kinetic and gravitational energies to the total mass, $K/M$ and $M_g/M$.  The relative contribution of  the kinetic energy to the total mass is defined as
\begin{equation}
\frac{K}{M}=\frac{1}{M}\left[\frac{f^2}{4\,m}\int d^3y\,\sqrt{\frac{B(y)}{A(y)}} Z'(y)^2\right].
\end{equation}
It may seem, after a cursory look at Table  \ref{data}, that the importance of the kinetic term is decreasing with decreasing $Z(0)$. In fact, it is easy to see that the kinetic term scales with $\Delta^2$, and so $K/ (M\, \Delta^2)$ hardly changes through the full range of solutions (see blue line of Figure \ref{KandGplot}).  In other words, $K\,/\,(M\,\Delta^2)$ is essentially constant in the interval $0.01\leq Z(0)\leq 10$. It only varies somewhat faster in the unphysical region $Z(0)>10$. On the transition branch, where the quadratic term of the self-interaction potential can be cleanly separated from the rest of the axion potential but the contribution of gravity remains small, the self-interaction term is also of $\Ocal(\Delta^2\,M)$, and so the kinetic term and the self-interaction term are equally important 
%throughout the dense and dilute regions, 
for solving the EKG equations.

Let us consider now the gravity term.  The gravity term is weak, of $\Ocal(\delta \,M)$, {\em throughout the dense and dilute regions}.  However, near the dilute maximum and along the dilute branch, where $\Delta^2 \lesssim \Ocal(\delta)$, the  gravitational energy becomes of similar magnitude with the kinetic and self-interaction energies; gravity thus plays an important part in determining the dilute maximum of mass and other properties of dilute solutions.  In fact, there would not be a dilute maximum of the mass spectrum without the contribution from gravity.  

In the dense region and in part of the transition branch, where $\delta\ll \Delta^2$, gravity plays a negligible role in solving the EKG equations.  Calculating the contribution of this term to the total mass of dense axion stars is more difficult than for other terms, because it contributes by a minuscule amount.  Though we performed all numerical integrations using the full set of equations, (\ref{EKG}), it is very difficult to use those calculations to give a direct estimate of the gravitational contribution for QCD parameters, since $\delta\simeq 2.5\times 10^{-15}$.

The easiest way to estimate the contribution is by expansion in the parameter $\delta\ll 1.$ One can expand Eqs. (\ref{EKG}) in a power series by defining $A=1+\delta\,a$ and $B=1+\delta\,b$. In leading order of $\delta$, the EKG equations take the form
\begin{align} \label{EKG2}
a'(y)&=-\frac{a(y)}{y} + 2\pi\,y\left\{\epsilon{}^2\,Z(y)^2+Z'(y)^2+4[1-J_0(Z)]\right\},\nn\\
b'(y)&=\frac{a(y)}{y} + 2\pi\,y\left\{\epsilon{}^2 \, Z(y)^2+Z'(y)^2-4[1-J_0(Z)]\right\},\nn\\
Z''(y)&+\frac{2}{y}Z'(y) +\epsilon^2\,Z(y)-2\,J_1[Z(y)]=0.
\end{align}
The resulting mass can be written in the form
\begin{align}
M&=M_0 + M_g,
\end{align}
where
\begin{align}
M_0 \equiv &\frac{f^2}{4\,m}\int d^3y\left\{\epsilon^2 \,Z(y)^2+ Z'(y)^2+4[1-J_0(Z)] \right\},
\nn \\
M_g\equiv &\frac{\delta \, f^2}{8\,m}\int d^3y\left\{a(y)\left[\epsilon^2\,Z(y)^2-Z'(y)^2+4[1-J_0(Z)]\right]\right.\nn\\
&\qquad \quad \left. + b(y)\left[-\epsilon^2\,Z(y)^2+Z'(y)^2+4[1-J_0(Z)]\right]\right\}.
\end{align}
We have defined $M_g$ as the total gravitational contribution to the mass at leading order in $\d$.  Since there are no small parameters in the expansion besides $\delta$, the equations of motion imply that $a(y)=\Ocal(1)$ and $b(y)=\Ocal(1)$. Consequently, the relative corrections to the mass functional are of $\mathcal{O}(\delta)$ and scale with $\delta$.  The red line of Figure \ref{KandGplot} shows that $M_g\,/\,(M\,\delta)$ is slowly varying in the interval $0.01\lesssim Z(0)\lesssim 10$.  For $Z(0)\gtrsim 10$, the EKG formalism becomes unreliable due to the assumption of particle number conservation.

An axion star can decay due to particle number non\textcolor{green}{-}conserving processes. The axion is described by a Hermitian scalar field, and therefore particle number is not a conserved quantity. The leading decay amplitude has the behavior $\exp[-f(\D)/\D]$ where $f(\D)$ is a slowly varying function of $\D$, with a finite limit as $\D\to0$ \cite{ESW}; a similar result is also well-known in the literature on oscillons \cite{MTY,Gleiser,HertzbergOscillon,Salmi,DecayCFT}.  If $\D \lesssim0.05$, the axion star is stable against decay during the lifetime of the universe; dilute QCD axion stars belong to this category. We observe in the top panel of Figure \ref{fig:methods} that the mass spectrum has a dense minimum at $\Delta_c\simeq 0.35$. Near the dense minimum there are two possible axion star configurations for a given particle number: one with $\Delta < \Delta_c$ which is structurally unstable due to the fact that it is at a local energy maximum; and the other with $\Delta> \Delta_c$ which has a large enough binding energy such that it is short lived due to decay. Therefore QCD axion stars in the dense branch cannot survive until the current epoch.  

\section{Other Methods} \label{sec:OtherMethods}

We emphasize that the EKG equations constitute a very accurate description of axion stars along the dilute, transition, and dense branches of solutions up to roughly $Z(0) \gtrsim 10$. Nonetheless, alternative descriptions proliferate, and in this section we will point out the relevant differences in order to determine where various descriptions are applicable. Importantly, the definition of the total mass as the volume integral of $T_{00}$ is modified across each method, as explained below. The other important parameters describing each solution are the radius $R_{99}$, the binding energy parameter $\D$, the chemical potential $\mu$, and the central field value $Z(0)$; we illustrate the relationships between these parameters in the bottom panel of Figure \ref{fig:RvsM_dilute_methods} and in all panels of Figure \ref{fig:methods}. In brief, the methods we consider in this work are shown in Table \ref{tab:methods}, and described in detail below.

\begin{table*}[t]
\hspace{-0.7cm}
\begin{tabular}{| c | c | c | }
\hline
 Method & Abbreviation & Brief Description \\
 \hline \hline
 Einstein-Klein-Gordon & EKG & Equations of motion for scalar field coupled to gravity \\
 Klein-Gordon & KG & No-gravity limit of EKG \\
 Generalized Ruffini-Bonazzola & GRB & KG equation including leading higher-harmonic corrections \\
 Gross-Pitaevskii-Poisson & GPP & Nonrelativistic and weak-gravity limit of EKG \\
Thomas-Fermi & TF & No-kinetic energy limit of GPP\\
 Classical EKG & CEKG & EKG equations with original cosine potential \\
 Sine-Gordon & SG & No-gravity limit of CEKG \\
 \hline
\end{tabular}
\caption{Brief description of the calculation methods analyzed in this work.  We emphasize that the TF approximation is invalid for any branch of axion star solutions if the leading self-interaction is attractive.}
 \label{tab:methods}
\end{table*}

\subsection{Direct Extension of EKG Equations}
\subsubsection{Klein-Gordon (KG)} \label{sec:NoGrav}

As we have already pointed out (and as shown in Table \ref{data}), on the dense branch and along most of the transition branch of solutions, gravity effectively decouples. In that case, we can set the metric functions $A = B = 1$, so that the EKG system (\ref{EKG}) reduces to
\begin{equation} \label{KGnograv}
 Z''(y) + \frac{2}{y}Z'(y) + \e^2\,Z(y) - 2\,J_1\left(Z(y)\right) = 0.
\end{equation}
Indeed, we also used Eq. (\ref{KGnograv}) to calculate the physical parameters at the same values listed in Table \ref{data} and found results which were essentially identical to those of Table \ref{data}. In this limit the total mass is given by Eq. (\ref{densepotential}) with $A = B = 1$. In the bottom panel of Figure \ref{fig:RvsM_dilute_methods} and in all panels of Figure \ref{fig:methods}, the cyan lines are direct calculations in the non-gravitating limit, and it is very clear that in the parameter space we consider, the results are exactly equivalent to those of the full EKG system.

\subsubsection{Generalized Ruffini-Bonazzola (GRB)} \label{sec:GRB}

The derivation of the EKG equations~\eqref{EKG} used the expansion of the field operator proposed by RB, given in Eq.~\eqref{phiRB}. In this formalism, the field is linear in creation and annihilation operators of the ground state, a parameterization that is exact in the limit of zero self-interactions for an appropriately chosen wavefunction $R(r)$ of eigenenergy $m\,\epsilon$. However, when the binding energy becomes large, self-interactions can excite higher-order modes of energy $k\,m\,\epsilon$ (where $k$ is a positive integer) whose wavefunctions $R_k(r)$ do not satisfy the same equations of motion. It is possible to calculate the backreaction of the higher-order excitations $R_{k>1}(r)$ on the leading mode $R_1(r)$ and thereby determine the effective axion star wavefunction. Doing so requires the extension of the RB operator of Eq.~\eqref{phiRB}, and so we have referred to this procedure as a Generalized RB (GRB) formalism.

%Though the KG equation is fully relativistic, the RB approach does not necessarily capture all relevant contributions to the field. In particular, on the dense branch of solutions, the equations (\ref{EKG}) become inadequate because of a breakdown of the single-harmonic approximation \cite{WilczekASt}. The RB operator (\ref{phiRB}), which is an exact solution to the non-interacting scalar field theory, is only approximate when self-interactions are included. Thus, it is necessary to generalize the RB operator to include higher-harmonic contributions of the form \cite{GRB}
The critical input for the GRB calculation is the extension of the RB field operator to include higher-order modes coupled to higher powers of creation and annihilation operators:
\begin{equation} \label{phiGRB}
 \phi(t,r) = \sum_{k=1}^{\infty} R_k(r) \left[(a_0)^k\,e^{-i\,k\,\e\,m\,t} + (a_0^\dagger)^k\,e^{i\,k\,\e\,m\,t}\right],
\end{equation}
which we refer to as the GRB field operator. In this framework, $R_1(r) = R(r)$ is the leading approximation, and higher-order contributions from $R_{k>1}$ can be organized as an perturbative expansion in the small parameter $\D$. This is possible because given the rescaling of the leading wavefunction component given in Eq. \eqref{rescale}, the equations of motion naturally require that the higher-order wavefunctions are suppressed by higher powers of $\D$ as
\begin{equation}
 R_k(r) = \frac{f}{2\,N^{k/2}} Z_k(y) = \frac{f}{2\,N^{k/2}} \Delta^k\,Y_k(x).
\end{equation}
The equations of motion for $Z_1$, $Z_3$, etc. can thus be solved perturbatively to obtain the total wavefunction.

The equation of motion for $Z_1$ is given, at $\Ocal(\D^5)$ in the GRB expansion, by \cite{GRB}
\begin{equation} \label{GRB}
 \D^2\,Z_1 = \nabla_y^2\,Z_1 + \frac{Z_1^3}{8} Z_1^3 - \frac{3\,Z_1^5}{512}.
\end{equation}
In \cite{GRB}, this equation was solved numerically for $Z_1$; the total mass can be calculated at this order in $\D$ using
\begin{equation}
 M = \frac{f^2}{4\,m} \int d^3y \left[\left(2 - \D^2\right)Z_1^2 + Z_1'^2 - \frac{Z_1^4}{16} + \frac{Z_1^6}{512} \right].
\end{equation}
Note that the central field value $Z(0)$ is not precisely equal to $Z_1(0)$ in GRB, because of higher-order corrections to the total wavefunction. In what follows, we merely take $Z(0)=Z_1(0)$ for easy comparison to the other methods; as explained in \cite{GRB}, the corrections from e.g. $Z_3(0)$ are suppressed by $\D^2<1$ and are negligible for our purposes.

The resulting masses and radii as determined in GRB are represented by the black curves in the bottom panel of Figure \ref{fig:RvsM_dilute_methods}.  We observe perfect agreement with the EKG results at small $Z(0) \lesssim 1$, but deviations appear near the dense crossover and along the dense branch. In particular, the dense minimum mass is found at $M_{\rm min} \approx 463 f^2/m$ in GRB \cite{GRB}, whereas in Section \ref{sec:Relativistic} we found $M_{\rm min} \approx 390 f^2/m$. Because GRB takes into account leading corrections from higher-harmonics, we believe it to be the more accurate method in this regime.

At large $Z(0)$ (which is also large $\D$), the GRB equation \eqref{GRB} no longer has solutions, just as we observed for EKG in Section \ref{sec:Relativistic}.  The cutoff for GRB is represented by the black dots in Figures \ref{fig:RvsM_dilute_methods} and \ref{fig:methods}. It is interesting that the large-$\D$ cutoff on the dense branch occurs at a smaller value $\D=0.69$ in GRB compared to $\D \approx 1$ in EKG (see top panel of Figure \ref{fig:methods}); nonetheless, such large values of $\D$ remain unphysical for the reasons outlined in Section \ref{sec:Relativistic}. It would be interesting to see how the large-$\D$ cutoff changes at even higher order in the GRB expansion, though this topic is beyond the scope of the present work.

A potential downside of the GRB formalism is that gravity has not been included. Other methods for determining relativistic corrections in real scalar field theory suffer from a similar limitation \cite{MTY,GuthRelativistic,BraatenRelativistic} (though see \cite{EMTWY} for some preliminary steps in this direction). Indeed, for the purposes of this section (describing the crossover from transition to dense branches of solutions), this does not constitute a serious limitation, as gravity is completely negligible over that range of solutions. However, for scalar fields without self-interactions, or whose self-interactions are strong and repulsive, it is possible to form bound states with large gravitational potentials. In those cases, a full description of relativistic corrections to axion stars would need to include post-Newtonian corrections to the gravitational potential.

\subsection{Nonrelativistic Limit} \label{sec:Nonrelativistic}
\subsubsection{Gross-Pitaevskii-Poisson (GPP)} \label{sec:GPP}

At leading order in weak gravity, the EKG system (\ref{EKG}) reduces to
\begin{equation} \label{KGP}
 Z''(y) + \frac{2}{y}Z'(y) + \e^2\,Z(y) - 2\,J_1\left(Z(y)\right) + \frac{2\,V_g(y)}{m} Z(y)= 0,
\end{equation}
where
\begin{equation} \label{PoissonZ}
 \left( \frac{\pd^2}{\pd y^2} + \frac{2}{y}\frac{\pd}{\pd y}\right) \frac{V_g(y)}{m} = 2\pi\,\d\,Z(y)^2
\end{equation}
with $\d = f^2/M_P^2$. Therefore $V_g$ satisfies the Poisson equation sourced by the scalar wavefunction $Z$.  We have already pointed out that when $Z(0) = \Ocal(1)$, gravity becomes extremely negligible in the KG equation; this fact is now made transparent by the suppression of the RHS of Eq. \eqref{PoissonZ} by the factor of $\d$.

We note that, in analogy to the other formalisms discussed, a no-gravity limit of the GPP formalism, namely the Gross-Pitaevskii (GP) formalism, would be exactly equivalent to the GPP formalism in the dense and transition regions since gravity is effectively decoupled along these branches.  However, we emphasize that for the dilute branch of solutions, gravity plays an important role in the stability of the condensate.  Along this branch, the GPP formalism well describes and the GP formalism fails to accurately describe the condensate.

To obtain the nonrelativistic limit of Eq. (\ref{KGP}), one must assume $1 - \e \ll 1$, i.e. that the chemical potential is small $m(1 - \e) \equiv -\mu \ll m$. In that case, $\e^2 \approx 1 + 2\m/m$, and we obtain
\begin{equation} \label{GPPBraaten}
 Z'' + \frac{2}{y}Z' = -\frac{2}{m}\left[\m - m\left(\frac{J_1(Z)}{Z} - \frac{1}{2}\right) + V_g\right] Z
\end{equation}
The system (\ref{PoissonZ}) and (\ref{GPPBraaten}) is the Gross-Pitaevskii equation coupled to Poisson gravity, here abbreviated as GPP, which is the most prominent approximation to the EKG equations. For clarity, we note that Eq. (\ref{GPPBraaten}) is equivalent to the standard GP equation used to analyze axion stars \cite{ChavanisMR,ELSW}. This is made transparent by identifying, as in \cite{Ansatz}, the relationship between $Z(y)$ and the standard Schr\"odinger wavefunction $\psi$: 
\begin{equation}
 Z = \sqrt{\frac{2\,\psi^*\psi}{m\,f^2}}.
\end{equation}
Then, using $\psi(t) \propto e^{-i\,\m\,t}$ in the single-harmonic limit \cite{Nambu}, we can rewrite (\ref{GPPBraaten}) as
\begin{align}
 i\,\dot{\psi} &= \left\{- \frac{\nabla^2}{2m} + V_g\right.  \nn \\
 		&\,\, \left.+ \frac{\partial}{\partial (\psi^*\psi)}\left[
 			m^2\,f^2\,\left(1 - J_0\left(\sqrt{\frac{2\psi^*\psi}{m\,f^2}}\right)\right) - \frac{m}{2}\psi^*\psi\right]\right\}\psi.
\end{align}
It was shown in \cite{Ansatz} that at leading order in the self-interaction, the GPP equations are exactly equivalent to the infrared ($\D \ll 1$) limit of the EKG equations (\ref{EKG}) which we have reproduced in Eq. (\ref{rescale}); either way, these equations are appropriate for dilute axion stars, but constitute a very bad approximation beyond the crossover to the dense branch of axion stars due to a breakdown of the nonrelativistic criterion. 

The GPP system was used in \cite{BraatenDense} to analyze the dense branch of axion star solutions. To calculate the total mass in this method, one must first determine the binding energy in the condensate, given by
\begin{align} \label{totE}
 E &= \frac{f^2}{4\,m}\int d^3y \bigg[Z'(y)^2 + V_g(y)\,Z(y)^2 \nn \\
 		&\hspace{1.5cm} +4 \left(1 - \frac{1}{4}Z(y)^2 - J_0\left(Z(y)\right)\right)\bigg].
\end{align}
Then the total mass is 
\begin{equation} \label{GPPmass}
 M = \left(1 + \frac{E}{m\,N}\right)\,m\,N
\end{equation}
with
\begin{equation} \label{GPPN}
 N \equiv \frac{f^2}{2\,m^2}\int d^3y \,Z(y)^2.
\end{equation}
We illustrate the total mass $M$ with the physical radius $R_{99}$ (bottom panel of Figure \ref{fig:RvsM_dilute_methods}) and binding energy parameter $\D$ (top panel of Figure \ref{fig:methods}), where the GPP results are given by the black dotted lines. In the very nonrelativistic region where $Z(0) \ll 1$, the results of GPP are equivalent to that of the EKG method. Near the dense crossover at $Z(0)=\Ocal(1)$, GPP starts to deviate and along the dense branch, shows very different behavior, due to a breakdown of the nonrelativistic criterion we have described. In Figure \ref{fig:methods}, we show $\D$ (middle panel) and the chemical potential $\m$ (bottom panel) as functions of the central field value $Z(0)$; clearly, as $Z(0)$ grows, the nonrelativistic GPP approach becomes increasingly suspect, and once $Z(0) \gtrsim 10$, one expects extremely large relativistic corrections. A recent work has formulated a perturbative method to take relativistic corrections into account using a GPP-like formalism \cite{GuthRelativistic}; for a $\phi^4$ potential, the results are equivalent to those of the GRB method described in Section \ref{sec:GRB}.

The nonrelativistic criteron $1-\e\ll1$, or equivalently $\D \ll 1$, gives rise to several important simplifications: the quantity $N$ in Eq. \eqref{GPPN} is easily identified by the (approximately conserved) total number of particles; $|E/m\,N| \ll 1$ is a small binding energy correction to the total mass; and the chemical potential is similarly small, $|\m/m| \ll 1$. However, near the crossover to the dense branch of solutions, corrections from special relativity become large, leading to a breakdown of this criteron. In particular, $\D = \Ocal(1)$ implies a large decay rate, violating the approximate $N$-conservation \cite{ESW,MTY,EMSW,EMTWY}. Further, comparing Eqs. \eqref{totE} and \eqref{GPPN}, it is clear that the binding energy per particle can be $\Ocal(1)$ at large $Z(0) \gtrsim 1$. Finally, $\m = -m(1 - \e) = \Ocal(-m)$ implies a very small amount of energy is required to create new particles in the condensate, violating number conservation in yet another way. This ultra-relativistic fluid is very different from the standard cold, nonrelativistic condensate assumed in the derivation of the GPP equations. There is no reason to believe that the GPP equations constitute a reasonable approximation in this regime.
 
%In that work, they further simplify the equation by assuming the contribution of the kinetic energy is small, which is the so-called Thomas-Fermi approximation, though it is now clear that such an approximation is inappropriate.

\subsubsection{Thomas-Fermi Approximation}

A related limit analyzed in \cite{BraatenDense} is the Thomas-Fermi (TF) approximation, where the kinetic energy is neglected compared to the gravitational and self-interaction potentials. The TF limit of Eq. (\ref{GPPBraaten}) is
\begin{equation} \label{TF}
 V_g = \m - m\left(\frac{J_1(Z)}{Z} - \frac{1}{2}\right).
\end{equation}
Then, using Eq. (\ref{PoissonZ}), one obtains a single equation for $Z$ of the form
\begin{equation}
 \left( \frac{\pd^2}{\pd y^2} + \frac{2}{y}\frac{\pd}{\pd y}\right)\left(\frac{J_1(Z)}{Z}\right) = 2\pi\,\d\,Z^2.
\end{equation}
Though this was used to analyze the dense branch of axion stars originally, it is now understood that (as we pointed out in Section \ref{sec:Relativistic}) the kinetic energy is a crucial contribution to the equations of motion at any value of $Z(0)$ yet considered, and so the TF approximation fails as a description of axion stars on any branch of solutions if attractive self-interactions are assumed \cite{WilczekASt,BraatenReview}.  However, this approximation is valid for appropriate boundary conditions if repulsive attractions are assumed \cite{Colpi}. We have included it here for completeness, but do not analyze it further.

\subsection{Classical Equations of Motion}
\subsubsection{Classical EKG (CEKG)}

The scalar field $\phi$ represents an operator in the original axion field theory. To derive the EKG equations \eqref{EKG}, we have taken expectation values of the stress-energy tensor and KG equation, a procedure that modifies the structure of the self-interaction potential. In particular, the original cosine potential of Eq. (\ref{Vphi}) is changed to a Bessel function $J_0$ in the Einstein equations, and $V'(\phi) \propto \sin\left(\phi/f\right)$ in the Klein-Gordon equation changes to $J_1$. One can in principle use the original trigonometric functions directly and solve the EKG system.  

Taking $Z = \sqrt{2}\,\phi/f$, the equations of motions are
\begin{widetext}
\begin{align} \label{CEKG}
\frac{A'(y)}{A(y)}&=\frac{1-A(y)}{y} + 2\pi\,\delta\,y\,A(y)\left\{\frac{\epsilon{}^2\,Z(y)^2}{B(y)}
			+ \frac{Z'(y)^2}{A(y)}+4[1-\cos(Z/\sqrt{2})]\right\}, \nn\\
\frac{B'(y)}{B(y)}&=\frac{A(y)-1}{y} + 2\pi\,\delta\,y\,A(y)\left\{\frac{\epsilon{}^2 \, Z(y)^2}{B(y)}
			+ \frac{Z'(y)^2}{A(y)}-4[1-\cos(Z/\sqrt{2})]\right\},\nn\\
Z''(y)&=-\left[\frac{2}{y}+\frac{B'(y)}{2\,B(y)}-\frac{A'(y)}{2\,A(y)}\right]Z'(y)
			- A(y)\left[\frac{\epsilon{}^2\,Z(y)}{B(y)} - \sqrt{2}\sin(Z/\sqrt{2})\right].
\end{align}
\end{widetext}
The normalization of the field $Z$ must be chosen such that if the self-interactions decouple (for example, at extremely small $Z(0)$), the total mass
\begin{align} \label{MCEKG}
 M &= \frac{f^2}{4 \,m}\int d^3y \,\sqrt{A(y) B(y)} \left[\frac{\e^2\,Z^2}{B(y)} 
 				+ \frac{Z'(y)^2}{A(y)} \right. \nn \\
					&\hspace{2cm} \left.+ 4\left(1-\cos(Z/\sqrt{2})\right)\right]
\end{align}
reduces to Eq. \eqref{densepotential}. We refer to this set of equations as the Classical EKG (CEKG) system; it is classical in the sense that it is obtained by neglecting the fact that the field $\phi$ is a quantum operator.
Importantly, solutions to the CEKG system must be limited to the range $0 < (Z(0)/\sqrt{2}) < 2\pi$, because the interaction potential has a shift symmetry that must be maintained.  This cutoff defines the red dots in Figures \ref{fig:RvsM_dilute_methods} and \ref{fig:methods}

The resulting CEKG mass vs. radius curve is given by the red solid line in the bottom panel of Figure \ref{fig:RvsM_dilute_methods}; because the results are identical to the non-gravitating limit, we postpone the discussion to the next section.

\subsubsection{Sine-Gordon (SG)}

The CEKG equations are interesting, in part, because the non-gravitating limit of the system is the Sine-Gordon (SG) equation:
\begin{equation}
 Z''(y) + \frac{2}{y}Z'(y) + \e^2\,Z(y) - \sqrt{2}\sin\left(Z(y)/\sqrt{2}\right) = 0.
\end{equation}
(The nonstandard factors of $\sqrt{2}$ in the equation arise only due to our normalization conventions.) This fully classical equation of motion has been analyzed extensively in oscillon literature \cite{Gleiser,HertzbergOscillon,Salmi}, and more recently in the context of dense axion stars by \cite{Hertzberg}. As before, the shift symmetry enforces $Z(0)<2\,\sqrt{2}\,\pi$. The total mass is given by Eq. \eqref{MCEKG} with $A=B=1$.

The mass $M$, radius $R_{99}$, binding energy parameter $\D$, and chemical potential per unit mass $\mu/m$, as determined in the SG formalism, are illustrated by the yellow dashed lines in the bottom panel of Figure \ref{fig:RvsM_dilute_methods} and in all panels of Figure \ref{fig:methods}, which are identical to the CEKG results (red lines). As pointed out in \cite{Hertzberg}, the dense branch as defined by the SG equation does not extend far beyond the crossover point, due to the shift symmetry requirement. However, we emphasize that the use of the SG equation does not capture the underlying axion field theory. The field $\phi$ is an operator and must be interpreted as acting on some state of the system, which for an axion star is usually taken as an $N$-particle condensate or as a coherent state; if this is not taken into account, it leads to the discrepancy in interpreting the dense branch of axion stars.

It is also important that even in the nonrelativistic region, there is a difference between the CEKG and EKG results. The reason can be seen by comparing the leading-order self-interaction term in Eqs. \eqref{EKG} and \eqref{CEKG}, which is relevant on the transition branch. The numerical factor on the $\phi^4$ interaction term is different due to the expansion of $\sin(Z/\sqrt{2})$ as compared to $J_1(Z)$. Such a small difference is difficult to notice unless one is directly comparing methods, as we have done here. Of course, in the limit of very dilute axion stars (away from the dilute maximum mass), either method will return comparable results because the self-interaction becomes less important compared to the gravitational force.

 \section{Conclusions} \label{sec:Conclusions}
  
In this work, we have taken a global view of QCD axion stars, analyzing the full range of input parameters for calculation and comparing results of different methods found in the literature. Axion stars have macroscopic properties that can be described by three branches of solutions: a \emph{dilute} branch, which is stable both structurally and against decay; a \emph{transition} branch, which is structurally unstable; and a \emph{dense} branch, which is structurally stable but unstable to fast decay to relativistic axions. These three branches can be described by the Einstein-Klein-Gordon (EKG) equations using a single input parameter, often taken either as the central value of the wavefunction $0 < Z(0) < \infty$, or the chemical potential $-m < \m < 0$.
  
The EKG equations describe axion stars extremely well along the dilute and transition branches of solutions; between these two branches, there is a well-known maximum mass of $M_{\rm max} = 10\,M_P\,f/m$. Near the crossover to the dense branch, corrections to the scalar wavefunction coming from backreaction of higher-energy modes become important, but can be taken into account using perturbative corrections to the EKG equations \cite{GRB}. The size of relativistic corrections is controlled by a parameter $\D<1$, and at $\Ocal(\D^5)$ the crossover point from transition to dense branches takes place at a minimum mass $M_{\rm min} = 390\,f^2/m$. At very large central field values $Z(0) \gtrsim 10$, the EKG equations (and even its known extensions) break down completely due to extremely large binding energy and rapid violation of number conservation. We emphasize that no known method is adequate to describe dense axion stars at large mass.

We have pointed out throughout this work that on the dense branch and along most of the transition branch of solutions, gravity is completely negligible.  We verify this by analyzing both the relative contributions of different terms in the EKG equations of motion, and by comparing directly the non-gravitating limit of the equations of motion to the original. If the dense branch extends to very large masses (a claim which we reiterate is not well-understood), then at some point gravity may reappear as a relevant binding force. A verification of this claim would require calculations on the dense branch at very large masses, which is at present not possible.  With our current knowledge, then, we note that for QCD axion stars, and in fact for boson stars composed of axion-like particles with $f \ll M_P$, there is no need for any general relativistic corrections in modeling these condensates along the full dilute and transition branches of solutions as well as along the dense branch of solutions for $Z(0) \lesssim 10$ where the formalism used in this study begins to break down. 

Aside from the EKG approach and its higher-harmonic extensions, various alternative approaches have been proposed in the literature to describe axion stars on the dense branch. We point out that these approaches fail as well at the largest values of $Z(0)$, due to breakdown of the assumptions on which they are based. In particular, the Gross-Pitaevskii-Poisson (GPP) equations are based on a nonrelativistic approximation of the EKG equations, and give spurious results for $Z(0) \gtrsim 4$; we have denoted this region as \emph{quasi-relativistic}, as the chemical potential $\m \lesssim -0.1\,m$ there. At even larger $Z(0)$, corresponding to even smaller (negative) $\m \lesssim - m/2$, the system is ultra-relativistic and the GPP equations are not applicable at all.

Another approach we have analyzed is the \emph{classical} approach, which ignores the expectation values of the axion field $\phi$ and uses the original $\cos(\phi/f)$ potential in the calculations. We point out that this approach gives spurious results even on the transition branch of solutions, due to a mismatch in the coefficient of the leading self-interaction term; this mismatch is exacerbated when higher-order self-interactions become relevant, as on the dense branch. The Classical EKG equations, in the non-gravitating limit, reduce to the Sine-Gordon equation often used in classical field analyses of oscillons. Such solutions must be truncated at small masses on the dense branch in order to enforce the periodicity of the potential; for this reason, the classical equations also fail as a description of dense axion stars.

Dense axion stars, if they were not highly unstable due to relativistic particle emission, could have extremely interesting phenomenological consequences due to their extremely large densities. In a theory of a complex scalar field, rather than a Hermitian field (like the QCD axion), there can exist a dense branch which is not unstable because particle number can be conserved. Such a theory is interesting and may warrant further study.

In addition to the numerical methods used throughout this paper, one may also utilize the variational method in describing solutions along the transition and dense branches.  This method, although less precise than numerical methods, can be used to gain a qualitative understanding and analytic control of the solutions along these branches, and can easily be used to analyze dynamic systems.  A paper on this subject is currently in preparation.

  \section*{Acknowledgements}
  
  We are grateful to H. Zhang for valuable discussions. The work of J.E. is supported by the Zuckerman STEM Leadership Program.  M.L. is supported by the Ford Predoctoral Fellowship and the Ashford Fellowship.  L.S. thanks the Department of Physics at the University of Cincinnati for financial support in the form of the Violet M. Diller Fellowship. L.C.R.W. thanks the Aspen Center for Physics, which is supported by National Science Foundation grant PHY-1607611, where some of the research was conducted.

\end{document}